\documentclass[graybox,natbib,nosecnum]{svmult}
\bibpunct{(}{)}{;}{a}{}{,} 

\pdfoutput=1   

\usepackage{mathptmx}       
\usepackage{helvet}         
\usepackage{courier}        
\usepackage{type1cm}        

\usepackage{makeidx}         
\usepackage{graphicx}        
\usepackage{multicol}        
\usepackage[bottom]{footmisc}
\usepackage[normalem]{ulem}	
\usepackage{hyperref}  
\usepackage{amssymb}

\usepackage{soul}   

\makeindex             

\begin{document}

\title*{On the Detection of Extrasolar Moons and Rings}
\author{Ren\'e Heller}
\institute{Ren\'e Heller  \at Max Planck Institute for Solar System Research, Justus-von-Liebig-Weg 3, 37077 G\"ottingen, Germany, \email{heller@mps.mpg.de}}
%
%
\maketitle

\vspace{-3.8cm}
Revised 3 August 2017
\vspace{0.8cm}

\abstract{
Since the discovery of a planet transiting its host star in the year 2000, thousands of additional exoplanets and exoplanet candidates have been detected, mostly by NASA's Kepler space telescope. Some of them are almost as small as the Earth's moon. As the solar system is teeming with moons, more than a hundred of which are in orbit around the eight local planets, and with all of the local giant planets showing complex ring systems, astronomers have naturally started to search for moons and rings around exoplanets in the past few years. We here discuss the principles of the observational methods that have been proposed to find moons and rings beyond the solar system and we review the first searches. Though no exomoon or exoring has been unequivocally validated so far, theoretical and technological requirements are now on the verge of being mature for such discoveries.
}

\section{Introduction -- Why bother about moons?}

The moons of the solar system planets have become invaluable tracers of the local planet formation and bombardment, e.g. for the Earth-Moon system \citep{1976LPI.....7..120C,Rufu2017} and for Mars and its tiny moons Phobos and Deimos \citep{Rosenblatt2016}. The composition of the Galilean moons constrains the temperature distribution in the accretion disk around Jupiter 4.5 billion years ago \citep{1974Icar...21..248P,2002AJ....124.3404C,2015A&A...579L...4H}. And while the major moons of Saturn, Uranus, and Neptune might have formed from circumplanetary tidal debris disks \citep{2012Sci...338.1196C}, Neptune's principal moon Triton has probably been caught during an encounter with a minor planet binary \citep{2006Natur.441..192A}. The orbital alignment of the Uranian moon systems suggests a ``collisional tilting scenario'' \citep{2012Icar..219..737M} and implies significant bombardment of the young Uranus. A combination of these observations constrains the migration of the giant planets \citep{2011A&A...536A..57D}, planet-planet encounters \citep{2014AJ....148...25D}, bombardment histories \citep{2001Icar..151..286L}, and the properties of the early circumsolar disk \citep{2014RSPTA.37230174J}. The Pluto-Charon system can be considered a planetary binary rather than a planet-moon system, since its center of mass is outside the radius of the primary, at about 1.7 Pluto radii. A giant impact origin of this system delivers important constraints on the characteristic frequency of large impacts in the Kuiper belt region \citep{2005Sci...307..546C}. Planetary rings consist of relatively small particles, from sub-grain-sized to boulder-sized, and they are indicators of moon formation and moon tidal/geophysical activity; see Enceladus around Saturn \citep{2006Sci...311.1416S}.

Exomoon and exoring discoveries can thus be expected to deliver information on exoplanet formation on a level that is fundamentally new and inaccessible through exoplanet observations alone. Yet, although thousands of planets have been found beyond the solar system, no natural satellite has been detected around any of them. So one obvious question to ask is: Where are they?

\section{Early Searches for Exomoons and Exorings}

So far, most of the searches for exomoons have been executed as piggyback science on projects with a different primary objective. To give just a few examples, \citet{2001ApJ...552..699B} used the exquisite photometry of the \textit{Hubble} Space Telescope (HST) to observe four transits of the hot Jupiter HD\,209458\,b in front of its host star. As the star has a particularly high apparent brightness and therefore delivers very high signal-to-noise transit light curves, these observations would have revealed the direct transits of slightly super-Earth-sized satellites ($\gtrsim~1.2\,R_\oplus$; $R_\oplus$ being the Earth's radius) around HD\,209458\,b if such a moon were present. Alternatively, the gravitational pull from any moon that is more massive than about $3\,M_\oplus$ ($M_\oplus$ being the Earth's mass) could have been detected as well. Yet, no evidence for such a large moon was found. \citet{2001ApJ...552..699B} also constrained the presence of rings around HD\,209458\,b, which must be either extremely edge-on (so they barely have an effect on the stellar brightness during transit) or they must be restricted to the inner $1.8$ planetary radii around HD\,209458\,b.

In a similar vein, \citet{2006ApJ...636..445C} found no evidence for moons or rings around the hot Saturn HD\,149026\,b, \citet{2007A&A...476.1347P} found no moons or rings around HD\,189733\,b, and \citet{2015A&A...583A..50S} rejected the exoring hypothesis for 51\,Peg\,b. \citet{2010MNRAS.407.2625M} used ground-based observations to search for moons around WASP-3b by studying the planet's transit timing variations (TTVs) and transit duration variations (TDVs). Yet, as TDVs remained undetectable in that system an exomoon scenario seems very unlikely to cause the observed TTVs. More recently, \citet{2015ApJ...814...81H} scanned a sample of 21 transiting planets observed with the Kepler space telescope \citep{2010Sci...327..977B} and found no conclusive evidence for ring signatures. Later, \citet{2017A&A...603A.115L} published a search for moons and rings around the warm Jupiter-sized exoplanet CoRoT-9\,b based on infrared (4.5\,$\mu$m) photometry obtained with the Spitzer space telescope during two transits in 2010 and 2011. Moons larger than 2.5 Earth radii were excluded at the $3\,\sigma$ confidence level, and large silicate-rich (alternatively: icy) rings with inclinations $\gtrsim13^\circ$ (alternatively: $\gtrsim3^\circ$) against the line of sight were also excluded.

The \textit{Hunt for Exomoons with Kepler} (HEK) project \citep{2012ApJ...750..115K}, the first dedicated survey targeting moons around extrasolar planets, is probably the best bet for a near-future exomoon detection. Their analysis combines TTV and TDV measurements of exoplanets with searches for direct photometric transit signatures of exomoons. The most recent summary of their Bayesian photodynamical modeling \citep{2011MNRAS.416..689K} of exomoon transits around a total of 57 Kepler Objects of Interest has been presented by \citet{2015ApJ...813...14K}. Other teams found unexplained TTVs in many transiting exoplanets from the Kepler mission \citep{2013A&A...553A..17S}, but without additional TDVs or direct photometric transits, a robust exomoon interpretation is impossible.

\subsection{Tentative Detections of Exomoons and Exorings}
\label{sec:tentative}

While a definite exomoon discovery remains to be announced, some tentative claims have already been presented in the literature. One of the first exomoon claims was put forward by \citet{2014ApJ...785..155B} based on the microlensing event MOA-2011-BLG-262. Their statistical analysis of the microlensing light curve, however, has a degenerate solution with two possible interpretations. It turns out that an interpretation invoking a $0.11^{+0.21}_{-0.06}\,M_\odot$ star with a $17^{+28}_{-10}\,M_\oplus$ planetary companion at $0.95^{+0.53}_{-0.19}$\,AU is a more reasonable explanation than the hypothetical $3.2\,M_{\rm Jup}$-mass free-floating planet with a $0.47\,M_\oplus$-mass moon at a separation of 0.13\,AU. Sadly, the sources of microlensing events cannot be followed up. As a consequence, no additional data can possibly be collected to confirm or reject the exomoon hypothesis of MOA-2011-BLG-262.

In the same year, \citet{2014ApJ...785L..30B} proposed that the observed asymmetry in the transit light curve of the hot Jupiter HD\,189733\,b might be caused by an opaque plasma torus around the planet, which could be fed by a tidally active natural companion around the planet (which is not visible in the transit light curve itself, in this scenario). But an independent validation has not been demonstrated.

Using a variation of the exoplanet transit method, \citet{2015ApJ...806...51H} presented the first evidence of an exomoon population in the Kepler data. The author used what he refers to as a superstack, a combination of light curves from thousands of transiting exoplanets and candidates, to create an average transit light curve from Kepler\,with a very low noise-to-signal level of about 1 part per million. This superstack of a light curve exhibits an additional transit-like signature to both sides of the averaged planetary transit, possibly caused by many exomoons that are hidden in the noise of the individual light curves of each exoplanet. The depth of this additional transit candidate feature corresponds to an effective moon radius of $2120_{-370}^{+330}$\,km, or about  0.8 Ganymede radii. Interestingly, this signal is much more pronounced in the superstack of planets with orbital periods larger than about 35\,d, whereas more close-in planets do not seem to show this exomoon-like feature. This finding is in agreement with considerations of the Hill stability of moons, which states that stellar gravitational perturbations may perturb the orbit of a moon around a close-in planet such that the moon will be ejected \citep{2006MNRAS.373.1227D}.

\begin{figure*}[t]
\begin{center}
\includegraphics[width=1.\linewidth]{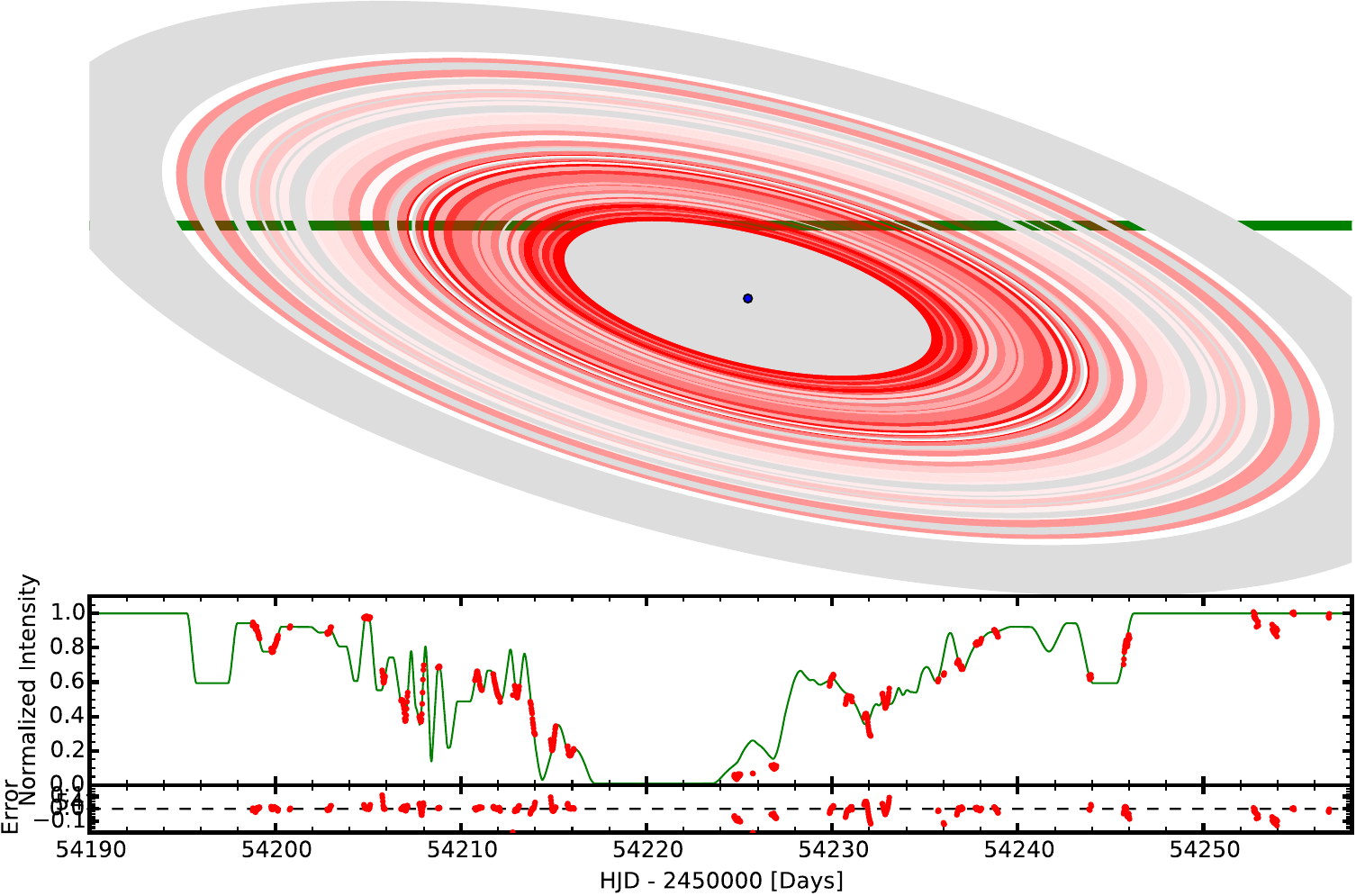}
\end{center}
\caption{The upper panel shows a computer model of a ring system transiting the star J1407 to explain the 70\,d of photometric observations from the Super Wide-Angle Search for Planets (SuperWASP) in the lower panel. The thick green line in the background of the upper panel represents the path of the star relative to the rings. Gray annuli indicate regions of the possible ring system that are not constrained by the data. Gradation of the red colors symbolizes the transmissivity of each ring. Note that the hypothesized central object of the ring system (possibly a giant planet) is not transiting the star. Image credit: \citet{2015ApJ...800..126K}. \copyright AAS. Reproduced with permission.}
\label{fig:J1407}
\end{figure*}

\citet{2017arXiv170708563T} also used a stacking method for 284 Kepler Objects of Interest with sufficiently high signal-to-noise ratios to constrain the occurrence rate of Galilean-analog moon systems to be $\eta<0.38$ with a 95\,\% confidence and $\eta=0.16_{-0.10}^{+0.13}$ with a 68.3\,\% confidence. If the signal in the stacked and phase-folded Kepler lightcurve is genuinely due to an exomoon population, then these moons would have average diameters roughly half the size of the Earth and would orbit their host planets at about 5 to 10 planetary radii.

Equally important, \citet{2017arXiv170708563T} present a viable exomoon candidate in orbit around Kepler-1625\,b. The satellite nature of what has provisionally been dubbed Kepler-1625\,b-i still needs to be confirmed (observations with the Hubble Space Telescope have been scheduled for October 2017), but the preliminary analysis suggests that this candidate would have a radius roughly the size of Neptune -- thus its informal designation as a ``Nept-moon'' -- and it would orbit a very massive Jupiter-sized planet. If validated, this binary would certainly present a benchmark system to study planet and moon formation because it would be fundamentally different from any planet-moon system found in the solar system.

Beyond the exquisite photometric data quality of the Kepler telescope, the COnvection ROtation and planetary Transits \citep[CoRoT;][]{2009A&A...506..411A}\,space mission also delivered highly accurate space-based stellar observations. One particularly interesting candidate object is CoRoT\,SRc01\,E2\,1066, which shows a peculiar bump near the center of the transit light curve that might be induced by the mutual eclipse of a transiting binary planet system \citep{2015ApJ...805...27L}, i.e., a giant planet with a very large and massive satellite. However, only one single transit of this object (or these two objects) has been observed, and so it is currently impossible to discriminate between a binary planet and a star-spot crossing interpretation of the data.

There has also been one supposed observation of a transiting ring system, which has been modeled to explain the curious brightness fluctuation of the 16\,Myr young K5 star 1SWASP\,J140747.93-394542.6 (J1407 for short) observed around 29 April 2007 \citep{2012AJ....143...72M}. The lower panel of Figure~\ref{fig:J1407} shows the observed stellar brightness variations, and the panel above displays the hypothesized ring system that could explain the data. This visualization nicely illustrates the connection between rings and moons, as the gaps in this proposed ring system could have been cleared by large moons that were caught in a stage of ongoing formation \citep{2015ApJ...800..126K}. The most critical aspect of this interpretation though is in the fact that the hypothesized central object has not been observed in transit. Another issue is that the orbital period of this putative ring system around J1407 can only be constrained to be {$\gtrsim 10$\,yr \citep[][and private communication with M.~Kenworthy]{2015MNRAS.446..411K}}. In other words, the periodic nature of this proposed transit event has not actually been established, and it could take decades to reobserve this phenomenon, if the interpretation were valid in the first place.

Almost all of the abovementioned studies have been inspired by the advent of space-based high-accuracy stellar photometric observations, mostly driven by the CoRoT and Kepler space telescopes. The newly gained access to this kind of a data quality has sparked huge interest in the possibility of novel detection methods for exomoons and exorings. In the following, we first consider detection methods for exomoons and then we discuss exoring detection methods.

\section{Detection Methods for Exomoons}

About a dozen different theoretical methods have been proposed to search and characterize exomoons. For the purpose of this review, we will group them into three classes: (1.) dynamical effects of the transiting host planet, (2.) direct photometric transits of exomoons, and (3) other methods.

\subsection{1. Dynamical Effects on Planetary Transits}
\label{sec:dynamical}

The moons of the solar system are small compared to their planet, and so the natural satellites of exoplanets are expected to be small as well. The depth ($d$) of an exomoon's photometric transit scales with the satellite radius ($R_{\rm s}$) squared: $d~{\propto}~R_{\rm s}^2$. Consequently, large exomoons could be relatively easy to detect (if they exist), but small satellites would tend to be hidden in the noise of the data. 

Alternatively, instead of hunting for the tiny brightness fluctuations caused by the moons themselves, it has been suggested that their presence could be derived indirectly by measuring the TTVs and TDVs of their host planets. The amplitudes of both quantities ($\Delta_{\rm TTV}$ and $\Delta_{\rm TDV}$) are linear in the mass of the satellite: $\Delta_{\rm TTV}~{\propto}~M_{\rm s}~{\propto}~\Delta_{\rm TTV}$ \citep{1999A&AS..134..553S,2009MNRAS.392..181K}. Hence, the dynamical effect of low-mass moons is less suppressed than the photometric effect of small-radius moons.

\subsubsection{Transit Timing Variation}

\begin{figure*}[t]
\begin{center}
\includegraphics[width=0.495\linewidth]{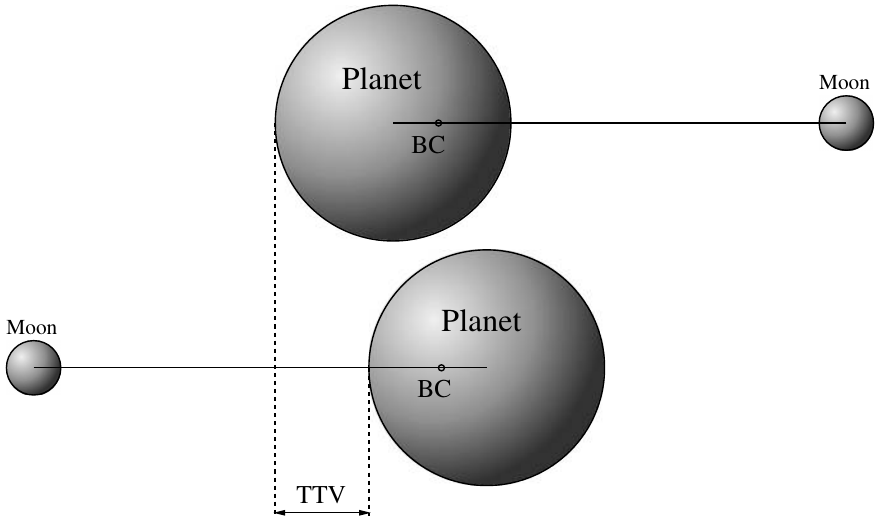}
\includegraphics[width=0.495\linewidth]{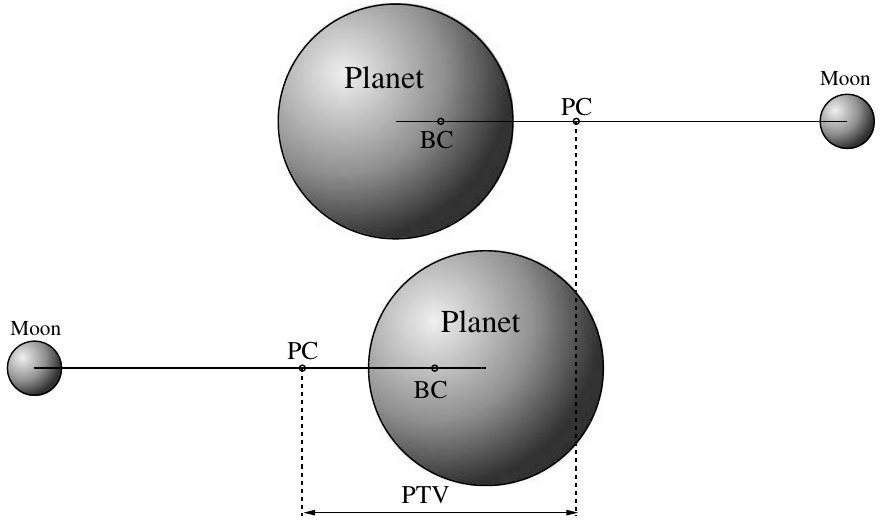}\vspace{.5cm} \\
\includegraphics[width=0.7\linewidth]{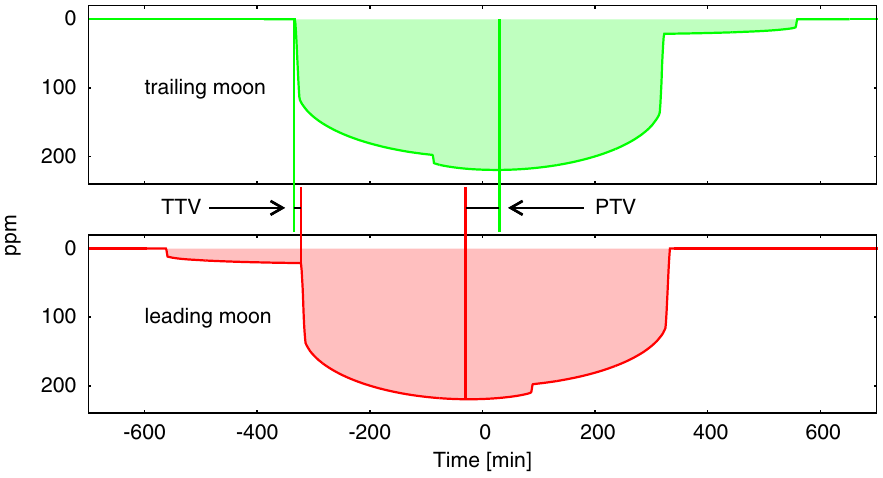}
\end{center}
\caption{Physical explanation of the barycentric TTV (upper left) and the photocentric TTV (upper right). The two light curves at the bottom illustrate both the TTV$_{\rm b}$ and the TTV$_{\rm p}$ (or PTV) for a moon that is trailing the planet (upper panel, green curve) and leading the planet during the stellar transit (lower panel, red curve). Image credit: \citet{2015PASP..127.1084S}. \copyright The Astronomical Society of the Pacific. Reproduced with permission. All rights reserved.}
\label{fig:TTVs}
\end{figure*}

In a somewhat simplistic picture, neglecting the orbital motion of a planet and its moon around their common center of gravity during their common stellar transit, TTVs are caused by the tangential offset of the planet from the planet-moon barycenter (see upper left illustration in Figure~\ref{fig:TTVs}, where ``BC'' denotes the barycenter). In a sequence of transits, the planet has different offsets during each individual event, assuming that it is not locked in a full-integer orbital resonance with its circumstellar orbit. Hence, its transits will not be precisely periodic but rather show TTVs, approximately on the order of seconds to minutes (compared to orbital periods of days to years).

Two flavors of observable TTV effects have been discussed in the literature. One is called the barycentric TTV method \citep[TTV$_{\rm b}$;][]{1999A&AS..134..553S,2009MNRAS.392..181K}, and one is referred to as the photocentric TTV method \citep[TTV$_{\rm p}$ or PTV;][]{2006A&A...450..395S,2007A&A...470..727S,2015PASP..127.1084S}. A graphical representation of both methods is shown in Figure~\ref{fig:TTVs}.

TTV$_{\rm b}$ measurements refer to the position of the planet relative to the planet-moon barycenter. From the perspective of a light curve analysis, this corresponds to measuring the time differences of the planetary transit only, e.g. of the ingress, center, and/or egress \citep{1999A&AS..134..553S}.

PTV measurements, on the other hand, take into account the photometric effects of both the planet and its moon, and so the corresponding amplitudes can actually be significantly larger than TTV$_{\rm b}$ amplitudes, details depending on the actual masses and radii of both objects \citep{2015PASP..127.1084S}.

\subsubsection{Transit Duration Variation}

Planetary TDVs can be caused by several effects. First, they can be produced by the change of the planet's tangential velocity component around the planet-moon barycenter between successive transits \citep[referred to as the TDV$_{\rm V}$ component;][]{2009MNRAS.392..181K}. When the velocity component in the planet-moon system that is tangential to the observer's line of sight adds to the circumstellar tangential velocity during the transit, then the event is relatively short. On the other hand, if the transit catches the planet during its reverse motion in the planet-moon system, then the total tangential velocity is lower than that of the barycenter, and so the planetary transit takes somewhat longer.

TDV effects can also be introduced if the planet-moon orbital plane is inclined with respect to the circumstellar orbital plane of their mutual center of gravity. In this case, the planet's apparent minimum distance from the stellar center will be different during successive transits, in more technical terms: its transit impact parameter will change between transits or, if the moon's orbital motion around the planet is fast enough, even during the transits. This can induce a TDV$_{\rm TIP}$ component in the transit duration measurements of the planet \citep{2009MNRAS.396.1797K}.

It is important to realize that the waveforms of the TTV and TDV curves are offset by an angle of $\pi/2$ \citep{2009MNRAS.392..181K}. In a more visual picture, when the TTV is zero, i.e. when the planet is along the line of sight with the planet-moon barycenter, then the corresponding TDV measurement is either largest (for moons on obverse motion) or smallest (for moons on reverse motion), since the planet would have the largest/smallest possible tangential velocity in the planet-moon binary system. This phase difference is key to breaking the degeneracy of simultaneous $M_{\rm s}$ and $a_{\rm s}$ measurements ($a_{\rm s}$ being the moon's semimajor axis around the planet). When plotted in a TTV-TDV diagram \citep{Montalto21122012,2013MNRAS.432.2549A}, the resulting ellipse contains predictable dynamical patterns, which can help to discriminate an exomoon interpretation of the data from a planetary perturber, and it may even allow the detection of multiple moons \citep{2016A&A...591A..67H}

\subsection{2. Direct Transit Signatures of Exomoons}
\label{sec:directtransit}

Like planets, moons could naturally imprint their own photometric transits into the stellar light curves, if they were large enough \citep{2011ApJ...743...97T}. The lower two panels of Figure~\ref{fig:TTVs} show an exomoon's contribution to the stellar bright variation in case the moon is trailing (upper light curve) or leading (lower light curve) its planet. Note that if the moon is leading, then its transit starts prior to the planetary transit, and so the exomoon transit affects the right part of the planetary transit in the light curve. As mentioned in section~``Dynamical Effects on Planetary Transits''\ref{sec:dynamical}, the key challenge is in the actual detection of this tiny contribution, which has hitherto remained hidden in the noise of exoplanet light curves.

As a variation of the transit method, it has been suggested that mutual planet-moon eclipses during their common stellar transit might betray the presence of an exomoon or binary planetary companion \citep{2009PASJ...61L..29S,2012MNRAS.420.1630P}. This is a particularly interesting method, since the mutual eclipses of two transiting planets have already been observed \citep{2016Natur.537...69D}. Yet, in the latter case, the two planets were known to exist prior to the observation of their common transit, whereas for a detection of an exomoon through mutual eclipses it would be necessary to test the data against a possible origin from star-spot crossings of the planet \citep{2015ApJ...805...27L} and to use an independent method for validation.

\subsubsection{Orbital Sampling Effect}
\label{sec:OSE}

\begin{figure*}[t]
\begin{center}
\includegraphics[width=0.7\linewidth]{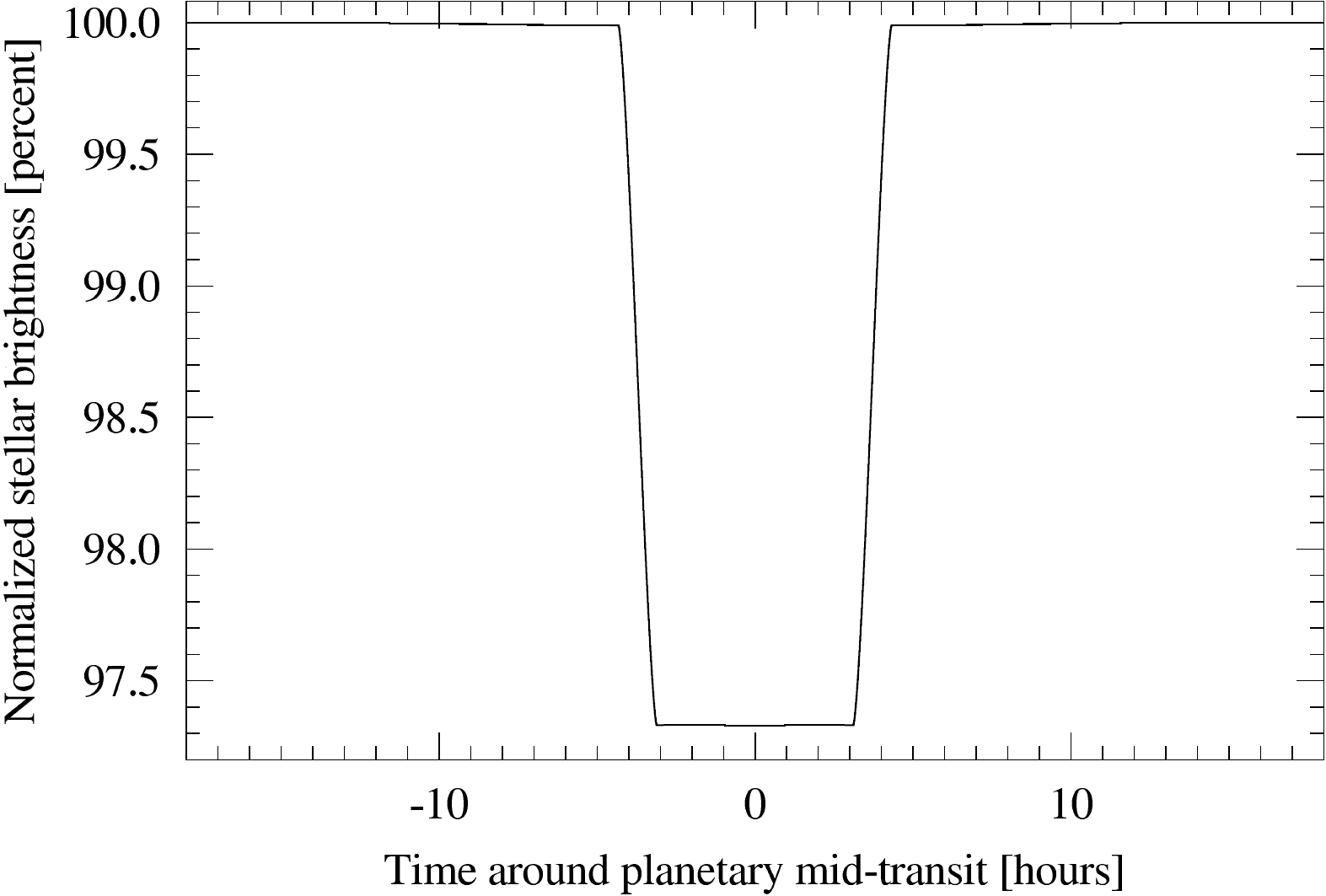}\vspace{.5cm} \\
\includegraphics[width=1.\linewidth]{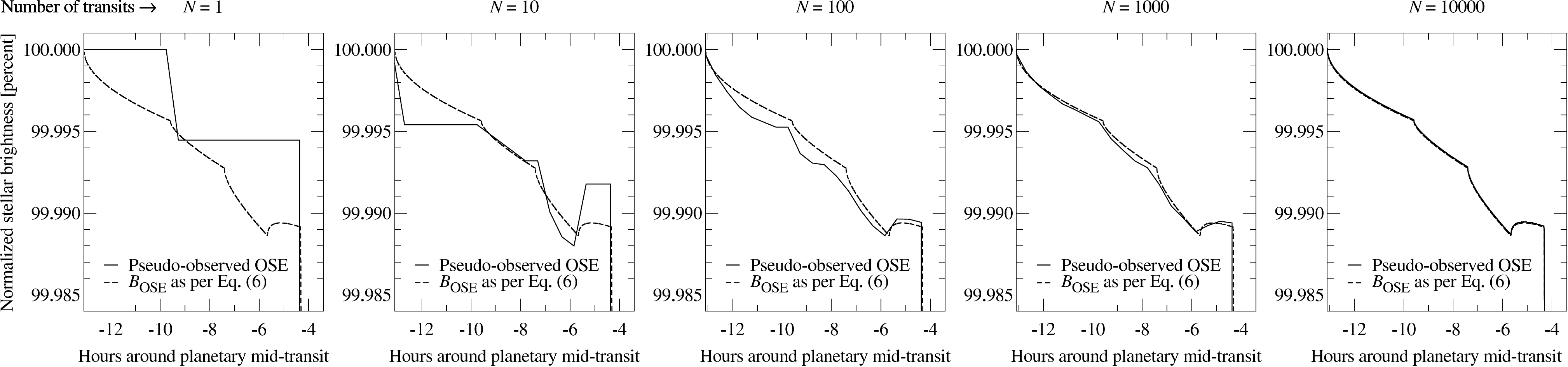}
\end{center}
\caption{Orbital sampling effect (OSE) of a simulated transiting exoplanet with moons. The upper panel shows a model of the phase-folded transit light curve of a Jupiter-sized planet around a $0.64\,R_\odot$ K dwarf star using an arbitrarily large number of transits. The planet is accompanied by three moons of $0.86\,R_\oplus$, $0.52\,R_\oplus$, and $0.62\,R_\oplus$ in radial size, but their contribution to the phase-folded light curve is barely visible with the naked eye. The lower row of panels shows a sequence of zooms into the prior-to-ingress part of the planetary transit. The evolution of the OSEs of the three moons is shown for an increasing number of transits ($N$) used to generate the phase-folded light curves. In each panel, the solid line shows the simulated phase-folded transit and the dashed line shows an analytical model, both curves assuming a star without limb darkening. Image credit: \citet{2014ApJ...787...14H}. \copyright AAS. Reproduced with permission.}
\label{fig:OSE}
\end{figure*}

One way to generate transit light curves with very high signal-to-noise ratios in order to reveal exomoons is by folding the measurements of several transits of the same object into one phase-folded transit light curve. Figure~\ref{fig:OSE} shows a simulation of this phase-folding technique, which is referred to as the orbital sampling effect \citep[OSE;][]{2014ApJ...787...14H,2016ApJ...820...88H}. The derived light curve does not effectively contain ``better'' data than the combination of the individual transit light curves (in fact it loses any information about the individual TTV and TDV measurements), but it enables astronomers to effectively search for moons in large data sets, as has been done by \citet{2015ApJ...806...51H} to generate superstack light curves from Kepler (see section~``Tentative Detections of Exomoons and Exorings''\ref{sec:tentative}).

\subsubsection{Scatter Peak}
\label{sec:scatter}

The minimum possible noise level of photometric light curves is given by the shot noise (or Poisson noise, white noise, time-uncorrelated noise), which depends on the number of photons collected and, thus, on the apparent brightness of the star. For the amount of photons typically collected with space-based optical telescopes, the minimum possible signal-to-noise ratio (SNR) of a light curve can be approximated as the square root of the number of photons ($n$): ${\rm SNR}~{\propto}~\sqrt{n}$. And so the SNR of a phase-folded transit light curve of a given planet goes down with the square root of the number of phase-folded transits ($N$): ${\rm SNR}_{\rm OSE}~{\propto}~\sqrt{N}$. In other words, for planets transiting photometrically quiet host stars, the noise-to-signal ratio (1/SNR) of the phase-folded light curve converges to zero for an increasing number of transits.

If the planet is accompanied by a moon, however, then the variable position of the moon with respect to the planet induces an additional noise component. As a consequence, and although the average light curve is converging toward analytical models (see Figure~\ref{fig:OSE}), the noise in the planetary transit is actually \textit{in}creasing due to the moon. For large $N$, once dozens and hundreds of transits can be phase-folded, the OSE becomes visible together with a peak in the noise, the latter of which has been termed the scatter peak \citep{2012MNRAS.419..164S}. As an aside, the superstack OSE candidate signal found by \citet{2015ApJ...806...51H} was not accompanied by any evidence of a scatter peak.

\subsection{3. Other Methods for Exomoon Detection}

In some cases, where the planet and its moon (or multiple moons) are sufficiently far from their host star, it could be possible to optically resolve the planet from the star. This has been achieved more than a dozen times now through a method known as direct imaging \citep{2008Sci...322.1348M}. Though direct imaging cannot, at the current stage of technology, deliver images of a resolved planet with individual moons, it might still be possible to detect the satellites. One could either try and detect the shadows and transits of the moons across their host planet in the integrated (i.e. unresolved) infrared light curve of the planet-moon system \citep{2007A&A...464.1133C,2016A&A...588A..34H}, or one could search for variations in the position of the planet-moon photocenter with respect to some reference object, e.g. another star or nearby exoplanet in the same system \citep{2007A&A...464.1133C,2015ApJ...812....5A}. Fluctuations in the infrared light received from the directly imaged planet $\beta$\,Pic\,b, as an example, could be due to an extremely tidally heated moon \citep{2013ApJ...769...98P} that is occasionally seen in transit or (not seen) during the secondary eclipse behind the planet. A related method is in the detection of a variation of the net polarization of light coming from a directly imaged planet, which might be caused by an exomoon transiting a luminous giant planet \citep{2016ApJ...824...76S}.

It could also be possible to detect exomoons through spectral analyses, e.g. via excess emission of giant exoplanets in the spectral region between 1 and 4\,$\mu$m \citep{2004AsBio...4..400W}, enhanced infrared emission by airless moons around terrestrial planets \citep{2009AsBio...9..269M,2011ApJ...741...51R}, and the stellar Rossiter-McLaughlin effect of a transiting planet with moons \citep{2010MNRAS.406.2038S,2012ApJ...758..111Z} or the Rossiter-McLaughlin effect of a moon crossing a directly imaged, luminous giant planet \citep{2014ApJ...796L...1H}.

Some more exotic exomoon detection methods invoke microlensing \citep{2002ApJ...580..490H,2010A&A...520A..68L,2014ApJ...785..155B,2014ApJ...785..156S}, pulsar timing variations \citep{2008ApJ...685L.153L}, modulations of radio emission from giant planets \citep{2014ApJ...791...25N,2016ApJ...821...97N}, or the generation of plasma tori around giant planets by volcanically active moons \citep{2014ApJ...785L..30B}.

\section{Detection Methods for Exorings}

Just like moons are very common around the solar system planets, rings appear to be a common feature as well. Naturally, the beautiful ring system around Saturn was the first to be discovered. Less obvious rings have also been detected around all other gas giants in the solar system and even around an asteroid \citep{2014Natur.508...72B}. In the advent of exoplanet detections, astronomers have thus started to develop methods for the detection of rings around planets outside the solar system.

\subsection{Direct Photometric Detection}

\begin{figure*}[t]
\includegraphics[width=1.0\linewidth]{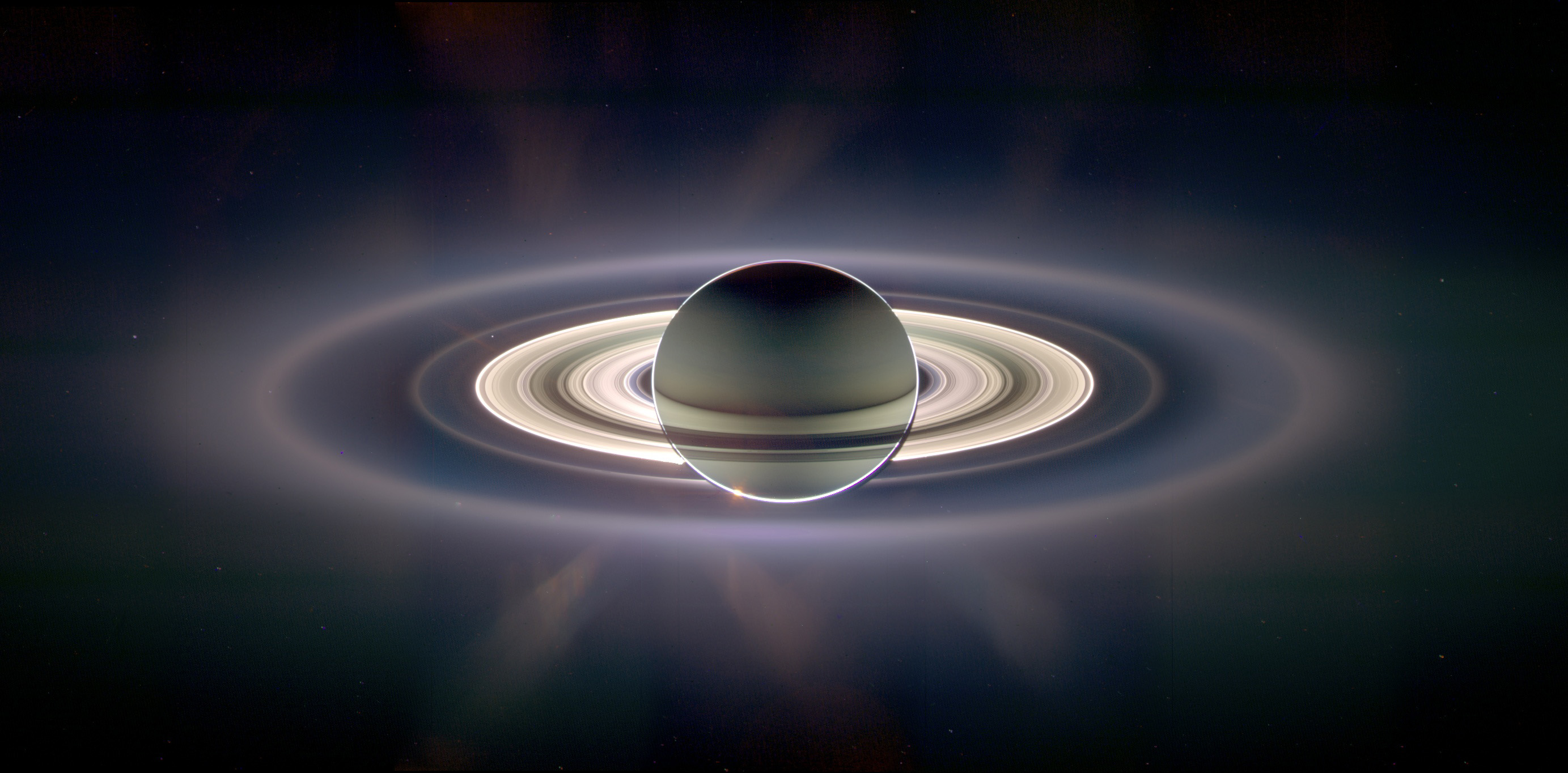}
\caption{Transit of Saturn and its ring system in front of the sun as seen by the Cassini spacecraft in September 2016. Note how the rings bend the sun light around the planet, an effect known as diffraction. Image credit: NASA/JPL/Space Science Institute.}
\label{fig:saturn}
\end{figure*}

The detection of rings around exoplanets is closely related to many of the abovementioned methods (see section~``Direct Transit Signatures of Exomoons''\ref{sec:directtransit}) of direct photometric transit observations of exomoons. Like moons, rings can cause additional dips in the planetary transit light curve \citep{2011ApJ...743...97T}. But rings do not induce any dynamical effects on the planet and, hence, there will be no TTVs or TDVs. In fact, a ring system can be expected to impose virtually the same pattern on each individual transit of its host planet because rings should look the same during each transit. Moons, however, would have a different position relative to the planet during individual transits (except for the case of full-integer orbital resonances between the circumstellar and the circumplanetary orbits). This static characteristic of the expected ring signals make it susceptible to misinterpretation, e.g. by a standard fit of a planet-only model to a hypothetically observed planet-with-ring light curve: in this case, the planet radius would be slightly overestimated, while the ring could remain undetected. However, the O-C diagram (O for observed, C for calculated) could still indicate the ring signature \citep{2004ApJ...616.1193B,2015ApJ...803L..14Z}

As a consequence, rings could induce a signal into the phase-folded transit light curve, which is very similar to the OSE (section~``Orbital Sampling Effect''\ref{sec:OSE}), since the latter is equivalent to a smearing of the moon over its circumplanetary orbit -- very much like a ring. The absence of dynamical effects like TTVs and TDVs, however, means that there would also be no scatter peak for rings (section~``Scatter Peak''\ref{sec:scatter}). Thus, an OSE-like signal in the phase-folded light curve without an additional scatter peak could indicate a ring rather than a moon.

One particular effect that has been predicted for light curves of transiting ring systems is diffraction, or forward-scattering \citep{2004ApJ...616.1193B}. Diffraction describes the ability of light to effectively bend around an obstacle, a property that is rooted in the waveform nature of light. In our context, the light of the host star encounters the ring particles along the line of sight, and those of which are ${\mu}$m- to 10\,m-sized will tend to scatter light into the forward direction, that is, toward the observer. In other words, rings cannot only obscure the stellar light during transit, they can also magnify it temporarily (see Figure~\ref{fig:saturn}). Moreover, rings can also reflect the stellar light to a significant degree, which can cause variations in the out-of-transit phases of the lightcurve \citep{2004A&A...420.1153A}.

\subsection{Other Detection Methods}

Beyond those potential ring signals in the photometric transit data, rings might also betray their presence in stellar transit spectroscopy. The crucial effect here is similar to the Rossiter-McLaughlin effect of transiting planets: as a transiting ring proceeds over the stellar disk, it modifies the apparent, disk-integrated radial velocity of its rotating host star. This is because the ring covers varying parts of the disk, all of which have a very distinct contribution to the rotational broadening of the stellar spectral lines \citep{2009ApJ...690....1O}. Qualitatively speaking, if the planet with rings transits the star in the same direction as the direction of stellar rotation, then the ring (and planet) will first cover the blue-shifted parts of the star. Hence, the stellar radial velocity will occur redshifted during about the first half of the transit -- and vice versa for the second half.

Another more indirect effect can be seen in the Fourier space (i.e. in the frequency domain rather than the time domain) of the transit light curve, where the ring can potentially stand out as an additional feature in the curve of the Fourier components as a function of frequency \citep{2015ApJ...807...65S}.

\section{Conclusions}

In this chapter, we discussed about a dozen methods that various researchers have worked out over little more than the past decade to search for moons and rings beyond the solar system. Although some of the original studies, in which these methods have been presented, expected that moons and rings could be detectable with the past CoRoT space mission or with the still active Kepler space telescope \citep{1999A&AS..134..553S,2004ApJ...616.1193B,2009MNRAS.400..398K,2014ApJ...787...14H}, no exomoon or exoring has been unequivocally discovered and confirmed as of today.

This is likely not because these extrasolar objects and structures do not exist, but because they are too small to be distinguished from the noise. Alternatively, and this is a more optimistic interpretation of the situation, those features could actually be detectable \textit{and} present in the available archival data (maybe even in the HST archival data of transiting exoplanets), but they just haven't been found yet. The absence of numerous, independent surveys for exomoons and exorings lends some credence to this latter interpretation: Out of the several thousands of exoplanets and exoplanet candidates discovered with the Kepler telescope alone, only a few dozen have been examined for moons and rings with statistical scrutiny \citep{2015ApJ...814...81H,2015ApJ...813...14K}.

It can be expected that the Kepler data will be fully analyzed for moons and rings within the next few years. Hence, a detection might still be possible. Alternatively, an independent, targeted search for moons/rings around planets transiting apparently bright stars -- e.g. using the HST, CHEOPS, or a 10\,m scale ground-based telescope, might deliver the first discoveries in the next decade. If none of these searches would be proposed or proposed but not granted, then it might take more than a decade for the PLAnetary Transits and Oscillations of stars (PLATO) mission \citep{2014ExA....38..249R} to find an exomoon or exoring in its large space-based survey of bright stars. Either way, it can be expected that exomoon and exoring discoveries will allow us a much deeper understanding of planetary systems than is possibly obtainable by planet observations alone.

\begin{acknowledgement}
This research has made use of NASA's Astrophysics Data System. The author thanks M.~Kenworthy, J.~Wambsganss, and J.~Schneider for helpful comments on an earlier version of this manuscript.
\end{acknowledgement}

\bibliographystyle{spbasicHBexo}  
\bibliography{heller} 

\begin{thebibliography}{79}
\providecommand{\natexlab}[1]{#1}
\providecommand{\url}[1]{{#1}}
\providecommand{\urlprefix}{URL }
\expandafter\ifx\csname urlstyle\endcsname\relax
  \providecommand{\doi}[1]{DOI~\discretionary{}{}{}#1}\else
  \providecommand{\doi}{DOI~\discretionary{}{}{}\begingroup
  \urlstyle{rm}\Url}\fi
\providecommand{\eprint}[2][]{\url{#2}}

\bibitem[{{Agnor} and {Hamilton}(2006)}]{2006Natur.441..192A}
{Agnor} CB {Hamilton} DP (2006) {Neptune's capture of its moon Triton in a
  binary-planet gravitational encounter}. \nat 441:192--194,
  \doi{10.1038/nature04792}

\bibitem[{{Agol} et~al(2015){Agol}, {Jansen}, {Lacy}, {Robinson}, and
  {Meadows}}]{2015ApJ...812....5A}
{Agol} E, {Jansen} T, {Lacy} B, {Robinson} T {Meadows} V (2015) {The Center of
  Light: Spectroastrometric Detection of Exomoons}. \apj 812:5,
  \doi{10.1088/0004-637X/812/1/5}, \eprint{1509.01615}

\bibitem[{{Arnold} and {Schneider}(2004)}]{2004A&A...420.1153A}
{Arnold} L {Schneider} J (2004) {The detectability of extrasolar planet
  surroundings. I. Reflected-light photometry of unresolved rings}. \aap
  420:1153--1162, \doi{10.1051/0004-6361:20035720}, \eprint{astro-ph/0403330}

\bibitem[{{Auvergne} et~al(2009){Auvergne}, {Bodin}, {Boisnard}, {Buey},
  {Chaintreuil}, {Epstein}, {Jouret}, {Lam-Trong}, {Levacher}, {Magnan},
  {Perez}, {Plasson}, {Plesseria}, {Peter}, {Steller}, {Tiph{\`e}ne}, {Baglin},
  {Agogu{\'e}}, {Appourchaux}, {Barbet}, {Beaufort}, {Bellenger}, {Berlin},
  {Bernardi}, {Blouin}, {Boumier}, {Bonneau}, {Briet}, {Butler}, {Cautain},
  {Chiavassa}, {Costes}, {Cuvilho}, {Cunha-Parro}, {de Oliveira Fialho},
  {Decaudin}, {Defise}, {Djalal}, {Docclo}, {Drummond}, {Dupuis}, {Exil},
  {Faur{\'e}}, {Gaboriaud}, {Gamet}, {Gavalda}, {Grolleau}, {Gueguen},
  {Guivarc'h}, {Guterman}, {Hasiba}, {Huntzinger}, {Hustaix}, {Imbert},
  {Jeanville}, {Johlander}, {Jorda}, {Journoud}, {Karioty}, {Kerjean},
  {Lafond}, {Lapeyrere}, {Landiech}, {Larqu{\'e}}, {Laudet}, {Le Merrer},
  {Leporati}, {Leruyet}, {Levieuge}, {Llebaria}, {Martin}, {Mazy}, {Mesnager},
  {Michel}, {Moalic}, {Monjoin}, {Naudet}, {Neukirchner}, {Nguyen-Kim},
  {Ollivier}, {Orcesi}, {Ottacher}, {Oulali}, {Parisot}, {Perruchot},
  {Piacentino}, {Pinheiro da Silva}, {Platzer}, {Pontet}, {Pradines},
  {Quentin}, {Rohbeck}, {Rolland}, {Rollenhagen}, {Romagnan}, {Russ}, {Samadi},
  {Schmidt}, {Schwartz}, {Sebbag}, {Smit}, {Sunter}, {Tello}, {Toulouse},
  {Ulmer}, {Vandermarcq}, {Vergnault}, {Wallner}, {Waultier}, and
  {Zanatta}}]{2009A&A...506..411A}
{Auvergne} M, {Bodin} P, {Boisnard} L et~al (2009) {The CoRoT satellite in
  flight: description and performance}. \aap 506:411--424,
  \doi{10.1051/0004-6361/200810860}, \eprint{0901.2206}

\bibitem[{{Awiphan} and {Kerins}(2013)}]{2013MNRAS.432.2549A}
{Awiphan} S {Kerins} E (2013) {The detectability of habitable exomoons with
  Kepler}. \mnras 432:2549--2561, \doi{10.1093/mnras/stt614},
  \eprint{1304.2925}

\bibitem[{{Barnes} and {Fortney}(2004)}]{2004ApJ...616.1193B}
{Barnes} JW {Fortney} JJ (2004) {Transit Detectability of Ring Systems around
  Extrasolar Giant Planets}. \apj 616:1193--1203, \doi{10.1086/425067},
  \eprint{astro-ph/0409506}

\bibitem[{{Ben-Jaffel} and {Ballester}(2014)}]{2014ApJ...785L..30B}
{Ben-Jaffel} L {Ballester} GE (2014) {Transit of Exomoon Plasma Tori: New
  Diagnosis}. \apjl 785:L30, \doi{10.1088/2041-8205/785/2/L30},
  \eprint{1404.1084}

\bibitem[{{Bennett} et~al(2014){Bennett}, {Batista}, {Bond}, {Bennett},
  {Suzuki}, {Beaulieu}, {Udalski}, {Donatowicz}, {Bozza}, {Abe}, {Botzler},
  {Freeman}, {Fukunaga}, {Fukui}, {Itow}, {Koshimoto}, {Ling}, {Masuda},
  {Matsubara}, {Muraki}, {Namba}, {Ohnishi}, {Rattenbury}, {Saito}, {Sullivan},
  {Sumi}, {Sweatman}, {Tristram}, {Tsurumi}, {Wada}, {Yock}, {MOA
  Collaboration}, {Albrow}, {Bachelet}, {Brillant}, {Caldwell}, {Cassan},
  {Cole}, {Corrales}, {Coutures}, {Dieters}, {Dominis Prester}, {Fouqu{\'e}},
  {Greenhill}, {Horne}, {Koo}, {Kubas}, {Marquette}, {Martin}, {Menzies},
  {Sahu}, {Wambsganss}, {Williams}, {Zub}, {PLANET Collaboration}, {Choi},
  {DePoy}, {Dong}, {Gaudi}, {Gould}, {Han}, {Henderson}, {McGregor}, {Lee},
  {Pogge}, {Shin}, {Yee}, {{$\mu$}FUN Collaboration}, {Szyma{\'n}ski},
  {Skowron}, {Poleski}, {Koz{\l}owski}, {Wyrzykowski}, {Kubiak},
  {Pietrukowicz}, {Pietrzy{\'n}ski}, {Soszy{\'n}ski}, {Ulaczyk}, {OGLE
  Collaboration}, {Tsapras}, {Street}, {Dominik}, {Bramich}, {Browne},
  {Hundertmark}, {Kains}, {Snodgrass}, {Steele}, {RoboNet Collaboration},
  {Dekany}, {Gonzalez}, {Heyrovsk{\'y}}, {Kandori}, {Kerins}, {Lucas},
  {Minniti}, {Nagayama}, {Rejkuba}, {Robin}, and {Saito}}]{2014ApJ...785..155B}
{Bennett} DP, {Batista} V, {Bond} IA et~al (2014) {MOA-2011-BLG-262Lb: A
  Sub-Earth-Mass Moon Orbiting a Gas Giant Primary or a High Velocity Planetary
  System in the Galactic Bulge}. \apj 785:155,
  \doi{10.1088/0004-637X/785/2/155}, \eprint{1312.3951}

\bibitem[{{Borucki} et~al(2010){Borucki}, {Koch}, {Basri}, {Batalha}, {Brown},
  {Caldwell}, {Caldwell}, {Christensen-Dalsgaard}, {Cochran}, {DeVore},
  {Dunham}, {Dupree}, {Gautier}, {Geary}, {Gilliland}, {Gould}, {Howell},
  {Jenkins}, {Kondo}, {Latham}, {Marcy}, {Meibom}, {Kjeldsen}, {Lissauer},
  {Monet}, {Morrison}, {Sasselov}, {Tarter}, {Boss}, {Brownlee}, {Owen},
  {Buzasi}, {Charbonneau}, {Doyle}, {Fortney}, {Ford}, {Holman}, {Seager},
  {Steffen}, {Welsh}, {Rowe}, {Anderson}, {Buchhave}, {Ciardi}, {Walkowicz},
  {Sherry}, {Horch}, {Isaacson}, {Everett}, {Fischer}, {Torres}, {Johnson},
  {Endl}, {MacQueen}, {Bryson}, {Dotson}, {Haas}, {Kolodziejczak}, {Van Cleve},
  {Chandrasekaran}, {Twicken}, {Quintana}, {Clarke}, {Allen}, {Li}, {Wu},
  {Tenenbaum}, {Verner}, {Bruhweiler}, {Barnes}, and
  {Prsa}}]{2010Sci...327..977B}
{Borucki} WJ, {Koch} D, {Basri} G et~al (2010) {Kepler Planet-Detection
  Mission: Introduction and First Results}. Science 327:977,
  \doi{10.1126/science.1185402}

\bibitem[{{Braga-Ribas} et~al(2014){Braga-Ribas}, {Sicardy}, {Ortiz},
  {Snodgrass}, {Roques}, {Vieira-Martins}, {Camargo}, {Assafin}, {Duffard},
  {Jehin}, {Pollock}, {Leiva}, {Emilio}, {Machado}, {Colazo}, {Lellouch},
  {Skottfelt}, {Gillon}, {Ligier}, {Maquet}, {Benedetti-Rossi}, {Gomes},
  {Kervella}, {Monteiro}, {Sfair}, {El Moutamid}, {Tancredi}, {Spagnotto},
  {Maury}, {Morales}, {Gil-Hutton}, {Roland}, {Ceretta}, {Gu}, {Wang},
  {Harps{\o}e}, {Rabus}, {Manfroid}, {Opitom}, {Vanzi}, {Mehret}, {Lorenzini},
  {Schneiter}, {Melia}, {Lecacheux}, {Colas}, {Vachier}, {Widemann},
  {Almenares}, {Sandness}, {Char}, {Perez}, {Lemos}, {Martinez},
  {J{\o}rgensen}, {Dominik}, {Roig}, {Reichart}, {Lacluyze}, {Haislip},
  {Ivarsen}, {Moore}, {Frank}, and {Lambas}}]{2014Natur.508...72B}
{Braga-Ribas} F, {Sicardy} B, {Ortiz} JL et~al (2014) {A ring system detected
  around the Centaur (10199) Chariklo}. \nat 508:72--75,
  \doi{10.1038/nature13155}, \eprint{1409.7259}

\bibitem[{{Brown} et~al(2001){Brown}, {Charbonneau}, {Gilliland}, {Noyes}, and
  {Burrows}}]{2001ApJ...552..699B}
{Brown} TM, {Charbonneau} D, {Gilliland} RL, {Noyes} RW {Burrows} A (2001)
  {Hubble Space Telescope Time-Series Photometry of the Transiting Planet of HD
  209458}. \apj 552:699--709, \doi{10.1086/320580}, \eprint{astro-ph/0101336}

\bibitem[{{Cabrera} and {Schneider}(2007)}]{2007A&A...464.1133C}
{Cabrera} J {Schneider} J (2007) {Detecting companions to extrasolar planets
  using mutual events}. \aap 464:1133--1138, \doi{10.1051/0004-6361:20066111},
  \eprint{astro-ph/0703609}

\bibitem[{{Cameron} and {Ward}(1976)}]{1976LPI.....7..120C}
{Cameron} AGW {Ward} WR (1976) {The Origin of the Moon}. In: Lunar and
  Planetary Science Conference, Lunar and Planetary Science Conference, vol~7

\bibitem[{{Canup}(2005)}]{2005Sci...307..546C}
{Canup} RM (2005) {A Giant Impact Origin of Pluto-Charon}. Science
  307:546--550, \doi{10.1126/science.1106818}

\bibitem[{{Canup} and {Ward}(2002)}]{2002AJ....124.3404C}
{Canup} RM {Ward} WR (2002) {Formation of the Galilean Satellites: Conditions
  of Accretion}. \aj 124:3404--3423, \doi{10.1086/344684}

\bibitem[{{Charbonneau} et~al(2006){Charbonneau}, {Winn}, {Latham}, {Bakos},
  {Falco}, {Holman}, {Noyes}, {Cs{\'a}k}, {Esquerdo}, {Everett}, and
  {O'Donovan}}]{2006ApJ...636..445C}
{Charbonneau} D, {Winn} JN, {Latham} DW et~al (2006) {Transit Photometry of the
  Core-dominated Planet HD 149026b}. \apj 636:445--452, \doi{10.1086/497959},
  \eprint{astro-ph/0508051}

\bibitem[{{Crida} and {Charnoz}(2012)}]{2012Sci...338.1196C}
{Crida} A {Charnoz} S (2012) {Formation of Regular Satellites from Ancient
  Massive Rings in the Solar System}. Science 338:1196,
  \doi{10.1126/science.1226477}, \eprint{1301.3808}

\bibitem[{{de Wit} et~al(2016){de Wit}, {Wakeford}, {Gillon}, {Lewis},
  {Valenti}, {Demory}, {Burgasser}, {Burdanov}, {Delrez}, {Jehin}, {Lederer},
  {Queloz}, {Triaud}, and {Van Grootel}}]{2016Natur.537...69D}
{de Wit} J, {Wakeford} HR, {Gillon} M et~al (2016) {A combined transmission
  spectrum of the Earth-sized exoplanets TRAPPIST-1 b and c}. \nat 537:69--72,
  \doi{10.1038/nature18641}, \eprint{1606.01103}

\bibitem[{{Deienno} et~al(2011){Deienno}, {Yokoyama}, {Nogueira}, {Callegari},
  and {Santos}}]{2011A&A...536A..57D}
{Deienno} R, {Yokoyama} T, {Nogueira} EC, {Callegari} N {Santos} MT (2011)
  {Effects of the planetary migration on some primordial satellites of the
  outer planets. I. Uranus' case}. \aap 536:A57,
  \doi{10.1051/0004-6361/201014862}

\bibitem[{{Deienno} et~al(2014){Deienno}, {Nesvorn{\'y}}, {Vokrouhlick{\'y}},
  and {Yokoyama}}]{2014AJ....148...25D}
{Deienno} R, {Nesvorn{\'y}} D, {Vokrouhlick{\'y}} D {Yokoyama} T (2014)
  {Orbital Perturbations of the Galilean Satellites during Planetary
  Encounters}. \aj 148:25, \doi{10.1088/0004-6256/148/2/25}, \eprint{1405.1880}

\bibitem[{{Domingos} et~al(2006){Domingos}, {Winter}, and
  {Yokoyama}}]{2006MNRAS.373.1227D}
{Domingos} RC, {Winter} OC {Yokoyama} T (2006) {Stable satellites around
  extrasolar giant planets}. \mnras 373:1227--1234,
  \doi{10.1111/j.1365-2966.2006.11104.x}

\bibitem[{{Han} and {Han}(2002)}]{2002ApJ...580..490H}
{Han} C {Han} W (2002) {On the Feasibility of Detecting Satellites of
  Extrasolar Planets via Microlensing}. \apj 580:490--493,
  \doi{10.1086/343082}, \eprint{astro-ph/0207372}

\bibitem[{{Heising} et~al(2015){Heising}, {Marcy}, and
  {Schlichting}}]{2015ApJ...814...81H}
{Heising} MZ, {Marcy} GW {Schlichting} HE (2015) {A Search for Ringed
  Exoplanets Using Kepler Photometry}. \apj 814:81,
  \doi{10.1088/0004-637X/814/1/81}, \eprint{1511.01083}

\bibitem[{{Heller}(2014)}]{2014ApJ...787...14H}
{Heller} R (2014) {Detecting Extrasolar Moons Akin to Solar System Satellites
  with an Orbital Sampling Effect}. \apj 787:14,
  \doi{10.1088/0004-637X/787/1/14}, \eprint{1403.5839}

\bibitem[{{Heller}(2016)}]{2016A&A...588A..34H}
{Heller} R (2016) {Transits of extrasolar moons around luminous giant planets}.
  \aap 588:A34, \doi{10.1051/0004-6361/201527496}, \eprint{1603.00174}

\bibitem[{{Heller} and {Albrecht}(2014)}]{2014ApJ...796L...1H}
{Heller} R {Albrecht} S (2014) {How to Determine an Exomoon's Sense of Orbital
  Motion}. \apjl 796:L1, \doi{10.1088/2041-8205/796/1/L1}, \eprint{1409.7245}

\bibitem[{{Heller} et~al(2015){Heller}, {Marleau}, and
  {Pudritz}}]{2015A&A...579L...4H}
{Heller} R, {Marleau} GD {Pudritz} RE (2015) {The formation of the Galilean
  moons and Titan in the Grand Tack scenario}. \aap 579:L4,
  \doi{10.1051/0004-6361/201526348}, \eprint{1506.01024}

\bibitem[{{Heller} et~al(2016{\natexlab{a}}){Heller}, {Hippke}, and
  {Jackson}}]{2016ApJ...820...88H}
{Heller} R, {Hippke} M {Jackson} B (2016{\natexlab{a}}) {Modeling the Orbital
  Sampling Effect of Extrasolar Moons}. \apj 820:88,
  \doi{10.3847/0004-637X/820/2/88}, \eprint{1603.07112}

\bibitem[{{Heller} et~al(2016{\natexlab{b}}){Heller}, {Hippke}, {Placek},
  {Angerhausen}, and {Agol}}]{2016A&A...591A..67H}
{Heller} R, {Hippke} M, {Placek} B, {Angerhausen} D {Agol} E
  (2016{\natexlab{b}}) {Predictable patterns in planetary transit timing
  variations and transit duration variations due to exomoons}. \aap 591:A67,
  \doi{10.1051/0004-6361/201628573}, \eprint{1604.05094}

\bibitem[{{Hippke}(2015)}]{2015ApJ...806...51H}
{Hippke} M (2015) {On the Detection of Exomoons: A Search in Kepler Data for
  the Orbital Sampling Effect and the Scatter Peak}. \apj 806:51,
  \doi{10.1088/0004-637X/806/1/51}, \eprint{1502.05033}

\bibitem[{{Jacobson} and {Morbidelli}(2014)}]{2014RSPTA.37230174J}
{Jacobson} SA {Morbidelli} A (2014) {Lunar and terrestrial planet formation in
  the Grand Tack scenario}. Philosophical Transactions of the Royal Society of
  London Series A 372:0174, \doi{10.1098/rsta.2013.0174}, \eprint{1406.2697}

\bibitem[{{Kenworthy} and {Mamajek}(2015)}]{2015ApJ...800..126K}
{Kenworthy} MA {Mamajek} EE (2015) {Modeling Giant Extrasolar Ring Systems in
  Eclipse and the Case of J1407b: Sculpting by Exomoons?} \apj 800:126,
  \doi{10.1088/0004-637X/800/2/126}, \eprint{1501.05652}

\bibitem[{{Kenworthy} et~al(2015){Kenworthy}, {Lacour}, {Kraus}, {Triaud},
  {Mamajek}, {Scott}, {S{\'e}gransan}, {Ireland}, {Hambsch}, {Reichart},
  {Haislip}, {LaCluyze}, {Moore}, and {Frank}}]{2015MNRAS.446..411K}
{Kenworthy} MA, {Lacour} S, {Kraus} A et~al (2015) {Mass and period limits on
  the ringed companion transiting the young star J1407}. \mnras 446:411--427,
  \doi{10.1093/mnras/stu2067}, \eprint{1410.6577}

\bibitem[{{Kipping}(2009{\natexlab{a}})}]{2009MNRAS.392..181K}
{Kipping} DM (2009{\natexlab{a}}) {Transit timing effects due to an exomoon}.
  \mnras 392:181--189, \doi{10.1111/j.1365-2966.2008.13999.x},
  \eprint{0810.2243}

\bibitem[{{Kipping}(2009{\natexlab{b}})}]{2009MNRAS.396.1797K}
{Kipping} DM (2009{\natexlab{b}}) {Transit timing effects due to an exomoon -
  II}. \mnras 396:1797--1804, \doi{10.1111/j.1365-2966.2009.14869.x},
  \eprint{0904.2565}

\bibitem[{{Kipping}(2011)}]{2011MNRAS.416..689K}
{Kipping} DM (2011) {LUNA: an algorithm for generating dynamic planet-moon
  transits}. \mnras 416:689--709, \doi{10.1111/j.1365-2966.2011.19086.x},
  \eprint{1105.3499}

\bibitem[{{Kipping} et~al(2009){Kipping}, {Fossey}, and
  {Campanella}}]{2009MNRAS.400..398K}
{Kipping} DM, {Fossey} SJ {Campanella} G (2009) {On the detectability of
  habitable exomoons with Kepler-class photometry}. \mnras 400:398--405,
  \doi{10.1111/j.1365-2966.2009.15472.x}, \eprint{0907.3909}

\bibitem[{{Kipping} et~al(2012){Kipping}, {Bakos}, {Buchhave}, {Nesvorn{\'y}},
  and {Schmitt}}]{2012ApJ...750..115K}
{Kipping} DM, {Bakos} G{\'A}, {Buchhave} L, {Nesvorn{\'y}} D {Schmitt} A (2012)
  {The Hunt for Exomoons with Kepler (HEK). I. Description of a New
  Observational project}. \apj 750:115, \doi{10.1088/0004-637X/750/2/115},
  \eprint{1201.0752}

\bibitem[{{Kipping} et~al(2015){Kipping}, {Schmitt}, {Huang}, {Torres},
  {Nesvorn{\'y}}, {Buchhave}, {Hartman}, and {Bakos}}]{2015ApJ...813...14K}
{Kipping} DM, {Schmitt} AR, {Huang} X et~al (2015) {The Hunt for Exomoons with
  Kepler (HEK): V. A Survey of 41 Planetary Candidates for Exomoons}. \apj
  813:14, \doi{10.1088/0004-637X/813/1/14}, \eprint{1503.05555}

\bibitem[{{Lecavelier des Etangs} et~al(2017){Lecavelier des Etangs},
  {H{\'e}brard}, {Blandin}, {Cassier}, {Deeg}, {Bonomo}, {Bouchy},
  {D{\'e}sert}, {Ehrenreich}, {Deleuil}, {D{\'{\i}}az}, {Moutou}, and
  {Vidal-Madjar}}]{2017A&A...603A.115L}
{Lecavelier des Etangs} A, {H{\'e}brard} G, {Blandin} S et~al (2017) {Search
  for rings and satellites around the exoplanet CoRoT-9b using Spitzer
  photometry}. \aap 603:A115, \doi{10.1051/0004-6361/201730554},
  \eprint{1705.01836}

\bibitem[{{Levison} et~al(2001){Levison}, {Dones}, {Chapman}, {Stern},
  {Duncan}, and {Zahnle}}]{2001Icar..151..286L}
{Levison} HF, {Dones} L, {Chapman} CR et~al (2001) {Could the Lunar ``Late
  Heavy Bombardment'' Have Been Triggered by the Formation of Uranus and
  Neptune?} \icarus 151:286--306, \doi{10.1006/icar.2001.6608}

\bibitem[{{Lewis} et~al(2008){Lewis}, {Sackett}, and
  {Mardling}}]{2008ApJ...685L.153L}
{Lewis} KM, {Sackett} PD {Mardling} RA (2008) {Possibility of Detecting Moons
  of Pulsar Planets through Time-of-Arrival Analysis}. \apjl 685:L153--L156,
  \doi{10.1086/592743}, \eprint{0805.4263}

\bibitem[{{Lewis} et~al(2015){Lewis}, {Ochiai}, {Nagasawa}, and
  {Ida}}]{2015ApJ...805...27L}
{Lewis} KM, {Ochiai} H, {Nagasawa} M {Ida} S (2015) {Extrasolar Binary Planets
  II: Detectability by Transit Observations}. \apj 805:27,
  \doi{10.1088/0004-637X/805/1/27}, \eprint{1504.06365}

\bibitem[{{Liebig} and {Wambsganss}(2010)}]{2010A&A...520A..68L}
{Liebig} C {Wambsganss} J (2010) {Detectability of extrasolar moons as
  gravitational microlenses}. \aap 520:A68, \doi{10.1051/0004-6361/200913844},
  \eprint{0912.2076}

\bibitem[{{Maciejewski} et~al(2010){Maciejewski}, {Dimitrov}, {Neuh{\"a}user},
  {Niedzielski}, {Raetz}, {Ginski}, {Adam}, {Marka}, {Moualla}, and
  {Mugrauer}}]{2010MNRAS.407.2625M}
{Maciejewski} G, {Dimitrov} D, {Neuh{\"a}user} R et~al (2010) {Transit timing
  variation in exoplanet WASP-3b}. \mnras 407:2625--2631,
  \doi{10.1111/j.1365-2966.2010.17099.x}, \eprint{1006.1348}

\bibitem[{{Mamajek} et~al(2012){Mamajek}, {Quillen}, {Pecaut}, {Moolekamp},
  {Scott}, {Kenworthy}, {Collier Cameron}, and {Parley}}]{2012AJ....143...72M}
{Mamajek} EE, {Quillen} AC, {Pecaut} MJ et~al (2012) {Planetary Construction
  Zones in Occultation: Discovery of an Extrasolar Ring System Transiting a
  Young Sun-like Star and Future Prospects for Detecting Eclipses by
  Circumsecondary and Circumplanetary Disks}. \aj 143:72,
  \doi{10.1088/0004-6256/143/3/72}, \eprint{1108.4070}

\bibitem[{{Marois} et~al(2008){Marois}, {Macintosh}, {Barman}, {Zuckerman},
  {Song}, {Patience}, {Lafreni{\`e}re}, and {Doyon}}]{2008Sci...322.1348M}
{Marois} C, {Macintosh} B, {Barman} T et~al (2008) {Direct Imaging of Multiple
  Planets Orbiting the Star HR 8799}. Science 322:1348,
  \doi{10.1126/science.1166585}, \eprint{0811.2606}

\bibitem[{Montalto et~al(2012)Montalto, Gregorio, Boué, Mortier, Boisse,
  Oshagh, Maturi, Figueira, Sousa, and Santos}]{Montalto21122012}
Montalto M, Gregorio J, Boué G et~al (2012) A new analysis of the wasp-3
  system: no evidence for an additional companion. Monthly Notices of the Royal
  Astronomical Society 427(4):2757--2771,
  \doi{10.1111/j.1365-2966.2012.21926.x},
  \urlprefix\url{http://mnras.oxfordjournals.org/content/427/4/2757.abstract},
  \eprint{http://mnras.oxfordjournals.org/content/427/4/2757.full.pdf+html}

\bibitem[{{Morbidelli} et~al(2012){Morbidelli}, {Tsiganis}, {Batygin}, {Crida},
  and {Gomes}}]{2012Icar..219..737M}
{Morbidelli} A, {Tsiganis} K, {Batygin} K, {Crida} A {Gomes} R (2012)
  {Explaining why the uranian satellites have equatorial prograde orbits
  despite the large planetary obliquity}. \icarus 219:737--740,
  \doi{10.1016/j.icarus.2012.03.025}, \eprint{1208.4685}

\bibitem[{{Moskovitz} et~al(2009){Moskovitz}, {Gaidos}, and
  {Williams}}]{2009AsBio...9..269M}
{Moskovitz} NA, {Gaidos} E {Williams} DM (2009) {The Effect of Lunarlike
  Satellites on the Orbital Infrared Light Curves of Earth-Analog Planets}.
  Astrobiology 9:269--277, \doi{10.1089/ast.2007.0209}, \eprint{0810.2069}

\bibitem[{{Noyola} et~al(2014){Noyola}, {Satyal}, and
  {Musielak}}]{2014ApJ...791...25N}
{Noyola} JP, {Satyal} S {Musielak} ZE (2014) {Detection of Exomoons through
  Observation of Radio Emissions}. \apj 791:25,
  \doi{10.1088/0004-637X/791/1/25}, \eprint{1308.4184}

\bibitem[{{Noyola} et~al(2016){Noyola}, {Satyal}, and
  {Musielak}}]{2016ApJ...821...97N}
{Noyola} JP, {Satyal} S {Musielak} ZE (2016) {On the Radio Detection of
  Multiple-exomoon Systems due to Plasma Torus Sharing}. \apj 821:97,
  \doi{10.3847/0004-637X/821/2/97}, \eprint{1603.01862}

\bibitem[{{Ohta} et~al(2009){Ohta}, {Taruya}, and {Suto}}]{2009ApJ...690....1O}
{Ohta} Y, {Taruya} A {Suto} Y (2009) {Predicting Photometric and Spectroscopic
  Signatures of Rings Around Transiting Extrasolar Planets}. \apj 690:1--12,
  \doi{10.1088/0004-637X/690/1/1}, \eprint{astro-ph/0611466}

\bibitem[{{P{\'a}l}(2012)}]{2012MNRAS.420.1630P}
{P{\'a}l} A (2012) {Light-curve modelling for mutual transits}. \mnras
  420:1630--1635, \doi{10.1111/j.1365-2966.2011.20151.x}, \eprint{1111.1741}

\bibitem[{{Peters} and {Turner}(2013)}]{2013ApJ...769...98P}
{Peters} MA {Turner} EL (2013) {On the Direct Imaging of Tidally Heated
  Exomoons}. \apj 769:98, \doi{10.1088/0004-637X/769/2/98}, \eprint{1209.4418}

\bibitem[{{Pollack} and {Reynolds}(1974)}]{1974Icar...21..248P}
{Pollack} JB {Reynolds} RT (1974) {Implications of Jupiter's Early Contraction
  History for the Composition of the Galilean Satellites}. \icarus 21:248--253,
  \doi{10.1016/0019-1035(74)90040-2}

\bibitem[{{Pont} et~al(2007){Pont}, {Gilliland}, {Moutou}, {Charbonneau},
  {Bouchy}, {Brown}, {Mayor}, {Queloz}, {Santos}, and
  {Udry}}]{2007A&A...476.1347P}
{Pont} F, {Gilliland} RL, {Moutou} C et~al (2007) {Hubble Space Telescope
  time-series photometry of the planetary transit of HD 189733: no moon, no
  rings, starspots}. \aap 476:1347--1355, \doi{10.1051/0004-6361:20078269},
  \eprint{0707.1940}

\bibitem[{{Rauer} et~al(2014){Rauer}, {Catala}, {Aerts}, {Appourchaux}, {Benz},
  {Brandeker}, {Christensen-Dalsgaard}, {Deleuil}, {Gizon}, {Goupil},
  {G{\"u}del}, {Janot-Pacheco}, {Mas-Hesse}, {Pagano}, {Piotto}, {Pollacco},
  {Santos}, {Smith}, {Su{\'a}rez}, {Szab{\'o}}, {Udry}, {Adibekyan}, {Alibert},
  {Almenara}, {Amaro-Seoane}, {Eiff}, {Asplund}, {Antonello}, {Barnes},
  {Baudin}, {Belkacem}, {Bergemann}, {Bihain}, {Birch}, {Bonfils}, {Boisse},
  {Bonomo}, {Borsa}, {Brand{\~a}o}, {Brocato}, {Brun}, {Burleigh}, {Burston},
  {Cabrera}, {Cassisi}, {Chaplin}, {Charpinet}, {Chiappini}, {Church},
  {Csizmadia}, {Cunha}, {Damasso}, {Davies}, {Deeg}, {D{\'{\i}}az}, {Dreizler},
  {Dreyer}, {Eggenberger}, {Ehrenreich}, {Eigm{\"u}ller}, {Erikson}, {Farmer},
  {Feltzing}, {de Oliveira Fialho}, {Figueira}, {Forveille}, {Fridlund},
  {Garc{\'{\i}}a}, {Giommi}, {Giuffrida}, {Godolt}, {Gomes da Silva},
  {Granzer}, {Grenfell}, {Grotsch-Noels}, {G{\"u}nther}, {Haswell}, {Hatzes},
  {H{\'e}brard}, {Hekker}, {Helled}, {Heng}, {Jenkins}, {Johansen},
  {Khodachenko}, {Kislyakova}, {Kley}, {Kolb}, {Krivova}, {Kupka}, {Lammer},
  {Lanza}, {Lebreton}, {Magrin}, {Marcos-Arenal}, {Marrese}, {Marques},
  {Martins}, {Mathis}, {Mathur}, {Messina}, {Miglio}, {Montalban}, {Montalto},
  {Monteiro}, {Moradi}, {Moravveji}, {Mordasini}, {Morel}, {Mortier},
  {Nascimbeni}, {Nelson}, {Nielsen}, {Noack}, {Norton}, {Ofir}, {Oshagh},
  {Ouazzani}, {P{\'a}pics}, {Parro}, {Petit}, {Plez}, {Poretti}, {Quirrenbach},
  {Ragazzoni}, {Raimondo}, {Rainer}, {Reese}, {Redmer}, {Reffert},
  {Rojas-Ayala}, {Roxburgh}, {Salmon}, {Santerne}, {Schneider}, {Schou},
  {Schuh}, {Schunker}, {Silva-Valio}, {Silvotti}, {Skillen}, {Snellen}, {Sohl},
  {Sousa}, {Sozzetti}, {Stello}, {Strassmeier}, {{\v S}vanda}, {Szab{\'o}},
  {Tkachenko}, {Valencia}, {Van Grootel}, {Vauclair}, {Ventura}, {Wagner},
  {Walton}, {Weingrill}, {Werner}, {Wheatley}, and
  {Zwintz}}]{2014ExA....38..249R}
{Rauer} H, {Catala} C, {Aerts} C et~al (2014) {The PLATO 2.0 mission}.
  Experimental Astronomy 38:249--330, \doi{10.1007/s10686-014-9383-4},
  \eprint{1310.0696}

\bibitem[{{Robinson}(2011)}]{2011ApJ...741...51R}
{Robinson} TD (2011) {Modeling the Infrared Spectrum of the Earth-Moon System:
  Implications for the Detection and Characterization of Earthlike Extrasolar
  Planets and Their Moonlike Companions}. \apj 741:51,
  \doi{10.1088/0004-637X/741/1/51}, \eprint{1110.3744}

\bibitem[{Rosenblatt et~al(2016)Rosenblatt, Charnoz, Dunseath, Terao-Dunseath,
  Trinh, Hyodo, Genda, and Toupin}]{Rosenblatt2016}
Rosenblatt P, Charnoz S, Dunseath KM et~al (2016) Accretion of phobos and
  deimos in an extended debris disc stirred by transient moons. Nature Geosci
  9(8):581--583, \urlprefix\url{http://dx.doi.org/10.1038/ngeo2742}

\bibitem[{Rufu et~al(2017)Rufu, Aharonson, and Perets}]{Rufu2017}
Rufu R, Aharonson O Perets HB (2017) A multiple-impact origin for the moon.
  Nature Geosci advance online publication:--,
  \urlprefix\url{http://dx.doi.org/10.1038/ngeo2866}

\bibitem[{{Samsing}(2015)}]{2015ApJ...807...65S}
{Samsing} J (2015) {Extracting Periodic Transit Signals from Noisy Light Curves
  using Fourier Series}. \apj 807:65, \doi{10.1088/0004-637X/807/1/65},
  \eprint{1503.03504}

\bibitem[{{Santos} et~al(2015){Santos}, {Martins}, {Bou{\'e}}, {Correia},
  {Oshagh}, {Figueira}, {Santerne}, {Sousa}, {Melo}, {Montalto}, {Boisse},
  {Ehrenreich}, {Lovis}, {Pepe}, {Udry}, and {Garcia
  Munoz}}]{2015A&A...583A..50S}
{Santos} NC, {Martins} JHC, {Bou{\'e}} G et~al (2015) {Detecting ring systems
  around exoplanets using high resolution spectroscopy: the case of 51 Pegasi
  b}. \aap 583:A50, \doi{10.1051/0004-6361/201526673}, \eprint{1509.00723}

\bibitem[{{Sartoretti} and {Schneider}(1999)}]{1999A&AS..134..553S}
{Sartoretti} P {Schneider} J (1999) {On the detection of satellites of
  extrasolar planets with the method of transits}. \aaps 134:553--560,
  \doi{10.1051/aas:1999148}

\bibitem[{{Sato} and {Asada}(2009)}]{2009PASJ...61L..29S}
{Sato} M {Asada} H (2009) {Effects of Mutual Transits by Extrasolar
  Planet-Companion Systems on Light Curves}. \pasj 61:L29, \eprint{0906.2590}

\bibitem[{{Sengupta} and {Marley}(2016)}]{2016ApJ...824...76S}
{Sengupta} S {Marley} MS (2016) {Detecting Exomoons around Self-luminous Giant
  Exoplanets through Polarization}. \apj 824:76,
  \doi{10.3847/0004-637X/824/2/76}, \eprint{1604.04773}

\bibitem[{{Simon} et~al(2007){Simon}, {Szatm{\'a}ry}, and
  {Szab{\'o}}}]{2007A&A...470..727S}
{Simon} A, {Szatm{\'a}ry} K {Szab{\'o}} GM (2007) {Determination of the size,
  mass, and density of ``exomoons'' from photometric transit timing
  variations}. \aap 470:727--731, \doi{10.1051/0004-6361:20066560},
  \eprint{0705.1046}

\bibitem[{{Simon} et~al(2015){Simon}, {Szab{\'o}}, {Kiss}, {Fortier}, and
  {Benz}}]{2015PASP..127.1084S}
{Simon} A, {Szab{\'o}} G, {Kiss} L, {Fortier} A {Benz} W (2015) {CHEOPS
  Performance for Exomoons: The Detectability of Exomoons by Using Optimal
  Decision Algorithm}. \pasp 127:1084--1095, \doi{10.1086/683392},
  \eprint{1508.00321}

\bibitem[{{Simon} et~al(2010){Simon}, {Szab{\'o}}, {Szatm{\'a}ry}, and
  {Kiss}}]{2010MNRAS.406.2038S}
{Simon} AE, {Szab{\'o}} GM, {Szatm{\'a}ry} K {Kiss} LL (2010) {Methods for
  exomoon characterization: combining transit photometry and the
  Rossiter-McLaughlin effect}. \mnras 406:2038--2046,
  \doi{10.1111/j.1365-2966.2010.16818.x}

\bibitem[{{Simon} et~al(2012){Simon}, {Szab{\'o}}, {Kiss}, and
  {Szatm{\'a}ry}}]{2012MNRAS.419..164S}
{Simon} AE, {Szab{\'o}} GM, {Kiss} LL {Szatm{\'a}ry} K (2012) {Signals of
  exomoons in averaged light curves of exoplanets}. \mnras 419:164--171,
  \doi{10.1111/j.1365-2966.2011.19682.x}, \eprint{1108.4557}

\bibitem[{{Skowron} et~al(2014){Skowron}, {Udalski}, {Szyma{\'n}ski}, {Kubiak},
  {Pietrzy{\'n}ski}, {Soszy{\'n}ski}, {Poleski}, {Ulaczyk}, {Pietrukowicz},
  {Koz{\l}owski}, {Wyrzykowski}, and {Gould}}]{2014ApJ...785..156S}
{Skowron} J, {Udalski} A, {Szyma{\'n}ski} MK et~al (2014) {New Method to
  Measure Proper Motions of Microlensed Sources: Application to Candidate
  Free-floating-planet Event MOA-2011-BLG-262}. \apj 785:156,
  \doi{10.1088/0004-637X/785/2/156}, \eprint{1312.7297}

\bibitem[{{Spahn} et~al(2006){Spahn}, {Schmidt}, {Albers}, {H{\"o}rning},
  {Makuch}, {Sei{\ss}}, {Kempf}, {Srama}, {Dikarev}, {Helfert},
  {Moragas-Klostermeyer}, {Krivov}, {Srem{\v c}evi{\'c}}, {Tuzzolino},
  {Economou}, and {Gr{\"u}n}}]{2006Sci...311.1416S}
{Spahn} F, {Schmidt} J, {Albers} N et~al (2006) {Cassini Dust Measurements at
  Enceladus and Implications for the Origin of the E Ring}. Science
  311:1416--1418, \doi{10.1126/science.1121375}

\bibitem[{{Szab{\'o}} et~al(2006){Szab{\'o}}, {Szatm{\'a}ry}, {Div{\'e}ki}, and
  {Simon}}]{2006A&A...450..395S}
{Szab{\'o}} GM, {Szatm{\'a}ry} K, {Div{\'e}ki} Z {Simon} A (2006) {Possibility
  of a photometric detection of ''exomoons''}. \aap 450:395--398,
  \doi{10.1051/0004-6361:20054555}, \eprint{astro-ph/0601186}

\bibitem[{{Szab{\'o}} et~al(2013){Szab{\'o}}, {Szab{\'o}}, {D{\'a}lya},
  {Simon}, {Hodos{\'a}n}, and {Kiss}}]{2013A&A...553A..17S}
{Szab{\'o}} R, {Szab{\'o}} GM, {D{\'a}lya} G et~al (2013) {Multiple planets or
  exomoons in Kepler hot Jupiter systems with transit timing variations?} \aap
  553:A17, \doi{10.1051/0004-6361/201220132}, \eprint{1207.7229}

\bibitem[{{Teachey} et~al(2017){Teachey}, {Kipping}, and
  {Schmitt}}]{2017arXiv170708563T}
{Teachey} A, {Kipping} DM {Schmitt} AR (2017) {HEK VI: On the Dearth of
  Galilean Analogs in Kepler and the Exomoon Candidate Kepler-1625b I}. ArXiv
  e-prints \eprint{1707.08563}

\bibitem[{{Tusnski} and {Valio}(2011)}]{2011ApJ...743...97T}
{Tusnski} LRM {Valio} A (2011) {Transit Model of Planets with Moon and Ring
  Systems}. \apj 743:97, \doi{10.1088/0004-637X/743/1/97}, \eprint{1111.5599}

\bibitem[{{Williams} and {Knacke}(2004)}]{2004AsBio...4..400W}
{Williams} DM {Knacke} RF (2004) {Looking for Planetary Moons in the Spectra of
  Distant Jupiters}. Astrobiology 4:400--403, \doi{10.1089/ast.2004.4.400}

\bibitem[{{Zhuang} et~al(2012){Zhuang}, {Gao}, and {Yu}}]{2012ApJ...758..111Z}
{Zhuang} Q, {Gao} X {Yu} Q (2012) {The Rossiter-McLaughlin Effect for Exomoons
  or Binary Planets}. \apj 758:111, \doi{10.1088/0004-637X/758/2/111},
  \eprint{1207.6966}

\bibitem[{{Zuluaga} et~al(2015){Zuluaga}, {Kipping}, {Sucerquia}, and
  {Alvarado}}]{2015ApJ...803L..14Z}
{Zuluaga} JI, {Kipping} DM, {Sucerquia} M {Alvarado} JA (2015) {A Novel Method
  for Identifying Exoplanetary Rings}. \apjl 803:L14,
  \doi{10.1088/2041-8205/803/1/L14}, \eprint{1502.07818}

\end{thebibliography}

\end{document}